\documentclass{PoS}

\usepackage{bm,amsmath,amssymb}
\usepackage{graphicx}

\title{Chiral symmetry restoration and deconfinement in strong magnetic fields}

\ShortTitle{Hot quark matter in strong magnetic fields}

\author{\speaker{Marco Ruggieri}
        \\
        Yukawa Institute for Theoretical Physics, Kyoto University, Kitashirakawa, Sakyoku, 606-8502 Kyoto, Japan\\
        E-mail: \email{ruggieri@yukawa.kyoto-u.ac.jp}}

\abstract{I review recent results obtained within chiral effective
models, on the phase structure of hot quark matter in a strong
magnetic background. After a brief introduction, I focus on the
results obtained within two chiral models improved with the
Polyakov loop. The models differ for the content of interactions,
but both of them are tuned to reproduce Lattice QCD thermodynamics
at zero and imaginary chemical potential. One of them takes into
account an explicit Polyakov loop dependence of the coupling; the
other one neglects this contribution, but takes into account
multi-quark interactions. A comparison between the phase diagrams
of the two models is presented.}

\FullConference{The many faces of QCD\\
        November 2-5, 2010\\
        Gent Belgium}

\begin{document}

\section{Introduction}
The study of the influence of external factors on the vacuum of
Quantum Chromodynamics (QCD), is one of the most attractive topics
of modern physics. Thanks to Lattice QCD simulations, it is
established that at vanishing quark chemical potential, two
crossovers take place in a narrow range of temperature; one for
quark deconfinement, and another one for the (approximate)
restoration of chiral
symmetry~\cite{deForcrand:2006pv,Aoki:2009sc,Bazavov:2009zn,Cheng:2009be,Karsch:2000kv}.
Besides, powerful analytic and semi-analytic techniques have been
developed to understand the coupling between chiral symmetry
restoration and deconfinement, see~\cite{Fischer:2009wc} and
references therein.

An alternative approach to the physics of strong interactions,
which is capable to capture some of the non-perturbative
properties of the QCD vacuum, is the Nambu-Jona Lasinio (NJL)
model~\cite{Nambu:1961tp}, see Refs.~\cite{revNJL} for reviews. In
this model, the QCD gluon-mediated interactions are replaced by
effective interactions among quarks, which are built in order to
respect the global symmetries of QCD. The NJL model has been
improved recently~\cite{Meisinger:1995ih,Fukushima:2003fw}, in
order to be capable to compute quantities which are related to the
confinement-deconfinement transition of QCD. In pure glue theory,
color confinement can be characterized by the expectation value of
the Polyakov loop~\cite{Polyakovetal}, which is an order parameter
for the center symmetry of the color gauge group. In theories with
dynamical fermions, the Polyakov loop is still a good indicator
for the confinement-deconfinement transition, as suggested by
Lattice data at zero chemical
potential~\cite{deForcrand:2006pv,Aoki:2009sc,Bazavov:2009zn,Cheng:2009be,Karsch:2000kv}.
Motivated by this property, the Polyakov extended Nambu-Jona
Lasinio model (P-NJL model) has been
introduced~\cite{Meisinger:1995ih,Fukushima:2003fw}, in which the
concept of {\em statistical confinement} replaces that of the true
confinement of QCD, and an effective potential describing
interaction among the chiral condensate and the Polyakov loop is
achieved by the coupling of quarks to a background temporal gluon
field, then integrating over quark fields in the functional
integral representation of the partition function. Chiral models
with the Polyakov loop have been studied extensively in many
contexts~\cite{Ratti:2005jh,Roessner:2006xn,Abuki:2008nm,Sakai:2008py,
Abuki:2008tx,Sakai:2010rp,Hell:2008cc,Herbst:2010rf,
Skokov:2010uh,Kahara:2008yg}.

In this talk, I report on results obtained in
Refs.~\cite{Fukushima:2010fe,Gatto:2010qs,Gatto:2010pt}, about
deconfinement and chiral symmetry restoration of hot quark matter
in a strong magnetic background. Related Lattice studies can be
found in~\cite{D'Elia:2010nq,Buividovich:2009my}; previous studies
of QCD and of QCD-like theories in a magnetic background can be
found in~\cite{Mizher:2010zb,Klevansky:1989vi,Chernodub:2010qx,
Frolov:2010wn}. Strong magnetic fields might be produced in
non-central heavy ion
collisions~\cite{Kharzeev:2007jp,Skokov:2009qp}. For example, at
the center-of-mass energy reachable at LHC, $\sqrt{s_{NN}} \approx
4.5$ TeV, the magnetic field can be as large as~\footnote{Here I
measure $eB$ in units of the vacuum squared pion mass $m_\pi^2$;
then, $eB=m_\pi^2$ corresponds to $B\approx 2\times 10^{14}$
Tesla.} $eB \approx 15 m_\pi^2$ according to~\cite{Skokov:2009qp}.
It has been argued that in these conditions, the sphaleron
transitions of finite temperature QCD, give rise to Chiral
Magnetic
Effect~\cite{Kharzeev:2007jp,Buividovich:2009wi,Fukushima:2008xe}.
This adds a phenomenological interest to the study presented here,
beside the purely theoretical one aforementioned.

The plan of the talk is as follow: after a brief introduction to
the model used in calculations, I discuss the results about
Polyakov loop expectation value and chiral condensate in a
magnetic background. Then, I collect these results in the form of
a phase diagram in the temperature-magnetic field strength plane.
Finally, I discuss briefly about improved results that take into
account a tree-level coupling among the chiral condensate and the
Polyakov loop, and show within the model how this coupling
modifies the response of hot quark matter to a magnetic
background.

\section{The model: propagator and one-loop thermodynamic potential}
I consider a model in which quark matter is described by the
following lagrangian density:
\begin{equation}
{\cal L} =\bar\psi\left(i\gamma^\mu D_\mu - m\right)\psi + {\cal
L}_I~;\label{eq:1ooo}
\end{equation}
here $\psi$ is the quark Dirac spinor in the fundamental
representation of the flavor $SU(2)$ and the color group. A sum
over color and flavor is understood. The covariant derivative
embeds the QED coupling of the quarks with the external magnetic
field, as well as the QCD coupling with the background gluon field
which is related to the Polyakov loop, see below. Furthermore,
\begin{equation}
{\cal L}_I = G\left[\left(\bar\psi\psi\right)^2 +
\left(i\bar\psi\gamma_5\bm\tau\psi\right)^2\right] +
G_8\left[\left(\bar\psi\psi\right)^2 +
\left(i\bar\psi\gamma_5\bm\tau\psi\right)^2\right]^2~.\label{eq:1}
\end{equation}
Here $\bm\tau$ correspond to the Pauli matrices in flavor space.
The 4-quark term corresponds to the original NJL
interaction~\cite{Nambu:1961tp}. On the other hand, the 8-quark
term was introduced in~\cite{Osipov:2005tq} to stabilize the NJL
vacuum. The interaction lagrangian is invariant under
$SU(2)_V\otimes SU(2)_A\otimes U(1)_V$. In the chiral limit, this
is the symmetry group of the action as well, if no magnetic field
is applied. However, this group is broken explicitly to
$U(1)_V^3\otimes U(1)_A^3\otimes U(1)_V$ if the magnetic field is
coupled to the quarks, because of the different electric charge of
$u$ and $d$ quarks. Here, the superscript $3$ in the $V$ and $A$
groups denotes the transformations generated by $\tau_3$,
$\tau_3\gamma_5$ respectively. Therefore, the chiral group in
presence of a magnetic field is $U(1)_V^3\otimes U(1)_A^3$. This
group is then explicitly broken by the quark mass term to
$U(1)_V^3$. I will present results obtained within a model with
$G$ taken as constant and $G_8 \neq 0$: this model is named
P-NJL$_8$ in~\cite{Sakai:2010rp}, and it is tuned in order to
reproduce Lattice data at zero and imaginary chemical potential.
Then I will compare these results with those of a slightly
different model, in which $G_8 = 0$ but $G$ depends on the
Polyakov loop. The latter model is named EPNJL
in~\cite{Sakai:2010rp}, and it is tuned in order to reproduce the
same Lattice data used to tune the P-NJL$_8$ model. Therefore,
both of them offer an effective description of the QCD
thermodynamics at zero and imaginary chemical potential. It is
thus interesting to compare their predictions on the response of
hot quark matter to a magnetic background.

To compute a temperature for the deconfinement crossover, I use
the expectation value of the Polyakov loop, $L$, defined as
\begin{equation}
L = \frac{1}{3}\text{Tr}_c\exp\left(i\beta\lambda_a A_4^a\right)~,
\end{equation}
where $A_4$ is a static, homogeneous and Euclidean background
temporal gluon field, coupled minimally to the quarks via the QCD
covariant derivative~\cite{Fukushima:2003fw}. In the Polyakov
gauge, which is convenient for this study, $A_0 = i\lambda_3 \phi
+ i \lambda_8 \phi^8$; moreover, since this work is done at zero
quark chemical potential, $L = L^\dagger$, which implies $A_4^8 =
0$. This choice is also motivated by the study
of~\cite{Mizher:2010zb}, where it is shown that the paramagnetic
contribution of the quarks to the thermodynamic potential induces
the breaking of the $Z_3$ symmetry, favoring the configurations
with a real-valued Polyakov loop .

I work in the one-loop approximation and neglect the pseudoscalar
condensates; moreover, I make the assumption that condensation
takes place only in the flavor channels $\tau_0$ and $\tau_3$. The
mean field interaction term Eq.~\eqref{eq:1} can be cast in the
form
\begin{equation}
{\cal L} = -2\bar\psi \psi\left[G \Sigma + 2G_8\Sigma^3\right] -G
\Sigma^2-3G_8\Sigma^4~,\label{eq:4}
\end{equation}
where $\Sigma = -\langle\bar{u}u + \bar{d}d\rangle$. The quark
propagator is easily expanded in terms of Landau levels using the
Leung-Ritus-Wang formalism~\cite{Ritus:1972ky}, namely
\begin{equation}
S_f(x,y) = \sum_{k=0}^\infty\int\frac{dp_0 dp_2 dp_3}{(2\pi)^4}
E_P(x)\Lambda_k \frac{1}{P\cdot\gamma - M}\bar{E}_P(y)~,
\label{eq:QP}
\end{equation}
where $E_P(x)$ corresponds to the eigenfunction of a charged
fermion in magnetic field, and $\bar{E}_P(x) \equiv
\gamma_0(E_P(x))^\dagger \gamma_0$. In the above equation $P =
(p_0 + i A_4,0,{\cal Q}\sqrt{2k|Q_f eB|},p_z)$, where $k
=0,1,2,\dots$ labels the $k^{\text{th}}$ Landau level, and ${\cal
Q} \equiv\text{sign}(Q_f)$, with $Q_f$ denoting the charge of the
flavor $f$; $\Lambda_k$ is a projector in Dirac space which keeps
into account the degeneracy of the Landau levels,
see~\cite{Ritus:1972ky} for details.

The trace of the $f$-quark propagator is minus the chiral
condensate $\langle\bar f f\rangle$, with $f=u,d$. Taking the
trace in coordinate and internal space, it is easy to show that
the following equation holds:
\begin{eqnarray}
\langle\bar f f\rangle &=& - N_c\frac{|Q_f
eB|}{2\pi}\sum_{k=0}^\infty \beta_k \int\frac{dp_z}{2\pi}
\frac{M_f}{\omega_f} {\cal C}(L,\bar L, T|p_z, k)~.\label{eq:CC}
\end{eqnarray}
Here,
\begin{eqnarray}
{\cal C}(L,\bar L, T|p_z, k) &=& U_\Lambda - 2{\cal N}(L,\bar L,
T|p_z, k)~,\label{eq:CCc}
\end{eqnarray}
and ${\cal N}$ denotes the {\em statistically confining} Fermi
distribution function,
\begin{eqnarray}
{\cal C}(L,\bar L, T|p_z, k) &=& \frac{1 + 2L e^{\beta\omega_f} +
Le^{2\beta\omega_f} }
  {1+3L e^{\beta\omega_f} + 3L e^{2\beta\omega_f} +
  e^{3\beta\omega_f}}~,
\end{eqnarray}
where
\begin{equation}
\omega_f^2 = p_z^2 + 2|Q_f e B|k + M^2~,~~~M =m_0 +2G\Sigma + 4
G_8 \Sigma^3~. \label{eq:MAsse}
\end{equation}
The coefficient $\beta_k = 2 -\delta_{k0}$ takes into account the
degeneracy of the Landau levels. The function $U_\Lambda$ is a
smooth UV-regulator, which is more suitable, from the numerical
point of view, in the present model calculation with respect to
the hard-cutoff which is used in analogous calculations without
magnetic field, see~\cite{Gatto:2010qs} for details.

In the one-loop approximation, the thermodynamic potential
$\Omega$ is given by
\begin{eqnarray}
\Omega &=& {\cal U}(L,\bar L,T) +U_M -\sum_{f=u,d}\frac{|Q_f
eB|}{2\pi}\sum_{k}\beta_k\int_{-\infty}^{+\infty}\frac{dp_z
}{2\pi} {\cal G}(L,\bar L,T|p_z,k)~.~\label{eq:OB}
\end{eqnarray}
In the above equation $U_M = G \Sigma^2 + 3 G_8 \Sigma^4$. The
last addendum arises from integration over the fermion
fluctuations in the functional integral representation of the
partition function; its contribution is written in terms of the
kernel
\begin{equation}
{\cal G}(L,\bar L, T|p_z,k) = N_c U_\Lambda(\bm p)\omega_f +
\frac{2}{\beta}\log{\cal F}~,
\end{equation}
with
\begin{equation}
{\cal F}(L,\bar L, T|p_z,k) = 1 +3L e^{-\beta\omega_f}+3\bar{L}
e^{-2\beta\omega_f}+e^{-3\beta\omega_f}~.
\end{equation}

The potential term $\mathcal{U}[L,\bar L,T]$ in Eq.~\eqref{eq:OB}
is built by hand in order to reproduce the pure gluonic lattice
data~\cite{Ratti:2005jh,Roessner:2006xn},
\begin{equation}
 \mathcal{U}[L,\bar L,T] = T^4\biggl\{-\frac{a(T)}{2}
  \bar L L + b(T)\ln\bigl[ 1-6\bar LL + 4(\bar L^3 + L^3)
  -3(\bar LL)^2 \bigr] \biggr\}~.
\label{eq:Poly}
\end{equation}
Analytic form of the coefficients $a(T)$, $b(T)$ can be found
in~\cite{Roessner:2006xn}. Here it is enough to notice that they
depend on one parameter, $T_0$, which corresponds to the
deconfinement scale. In the pure glue theory, $T_0 = 270$
\text{MeV}. Dynamical fermions induce a dependence of this
parameter on the number of active flavors~\cite{Herbst:2010rf}.
For the case of two light flavors to which I am interested here,
$T_0 = 212$ MeV.

\section{Deconfinement and chiral symmetry restoration}
\begin{figure}
\begin{center}
\includegraphics[width=7cm]{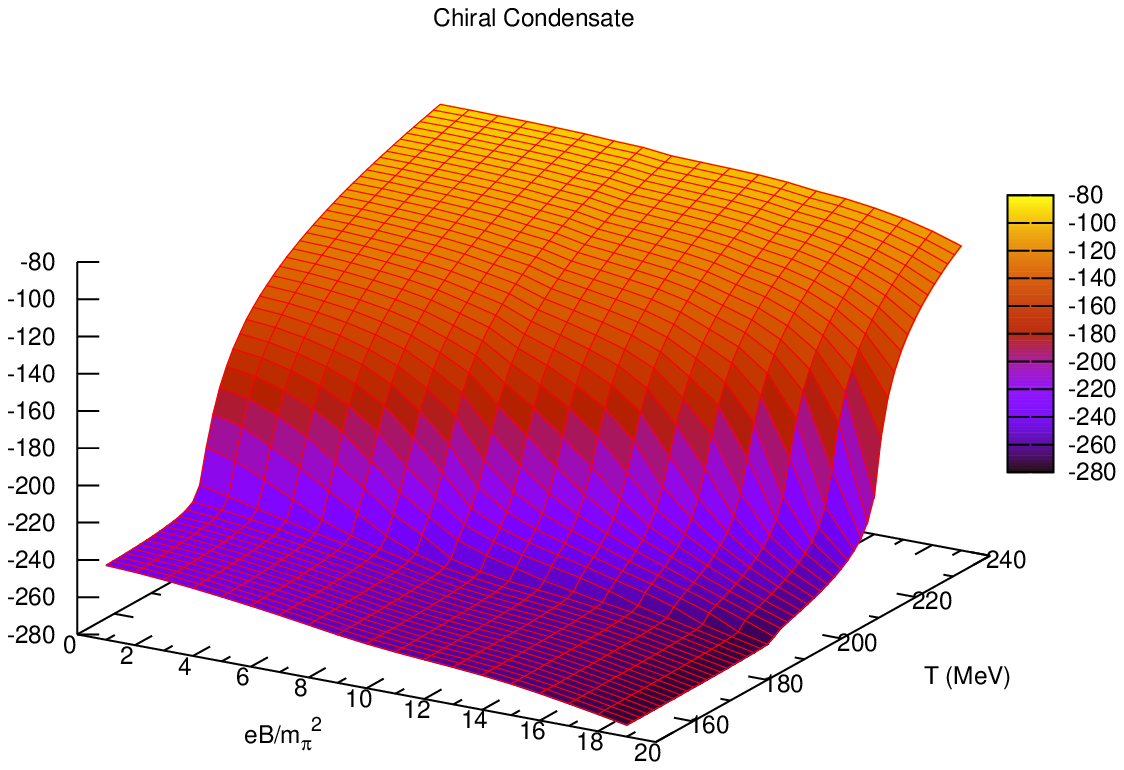}~~~\includegraphics[width=7cm]{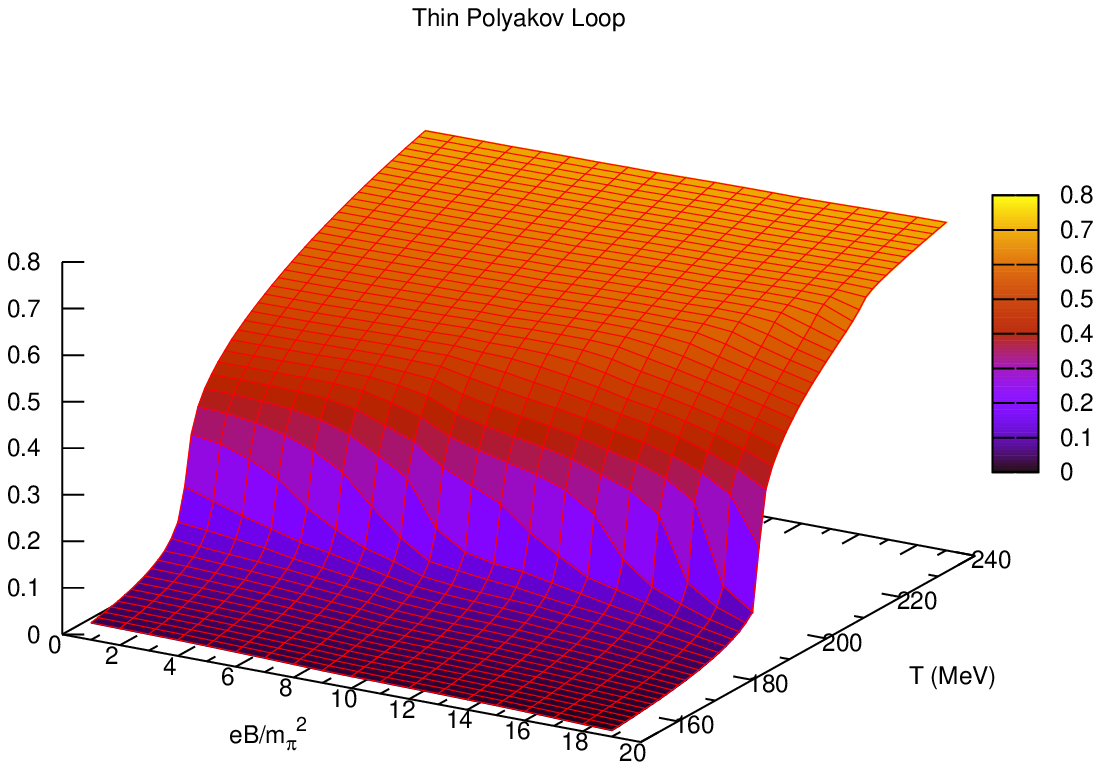}
\end{center}
\caption{\label{Fig:1} {\em Left panel:} Cubic root of the chiral
condensate, as a function of temperature and magnetic field. {\em
Right panel:} Expectation value of the Polyakov loop, $L$, as a
function of temperature and magnetic field strength. Results
correspond to the P-NJL$_8$ model.}
\end{figure}

In this Section, I firstly show the results obtained
in~\cite{Gatto:2010qs} within the model with interaction specified
in Eq.~\eqref{eq:1}. The numerical values of the parameters can be
found in the original reference. The model is named P-NJL$_8$,
following the nomenclature of~\cite{Sakai:2010rp}. Numerical data
for the chiral condensate and the Polyakov loop expectation value
are shown in Fig.~\ref{Fig:1}. They are obtained from the
minimization procedure of the one-loop thermodynamic potential in
Eq.~\eqref{eq:OB}.

Data on the chiral condensate, left panel of Fig.~\ref{Fig:1}, are
in agreement with the magnetic catalysis scenario: both the
numerical value of the chiral condensate and the critical
temperature, $T_\chi$, are increased by the magnetic field. The
latter is defined as the value which maximizes $|d \Sigma/ d T|$.
On the other hand, deconfinement crossover is poorly affected by
the magnetic field. The deconfinement temperature, $T_L$, is
identified with the temperature at which $dL/dT$ is maximum. At
$\bm B = 0$ the two crossovers take place simultaneously at
$T=175$ MeV. On the other hand, at $eB = 19 m_\pi^2$ the two
critical temperatures are split of an amount of $\approx 15\%$.
This split is also measured within a different chiral model
calculation~\cite{Mizher:2010zb}, and persists even in the chiral
limit~\cite{Fukushima:2010fe}.

I briefly compare the results of the P-NJL$_8$ model, with those
obtained within the EPNJL model~\cite{Sakai:2010rp} in which $G_8
= 0$, but an explicit dependence of the NJL coupling on the
Polyakov loop is taken into account (I take $L = L^\dagger$):
\begin{equation}
G = g\left(1 - \alpha_1 L^2 -2\alpha_2 L^3\right)~.\label{eq:Run}
\end{equation}
It has been shown in~\cite{Kondo:2010ts} that this kind of
dependence naturally arises in the low energy limit of QCD, see
also~\cite{Frasca:2008zp}. The numerical values of $\alpha_1$ and
$\alpha_2$ have been fixed in~\cite{Sakai:2010rp} by a best fit of
the available Lattice data at zero and imaginary chemical
potential of Ref.~\cite{D'Elia:2009qz}, $\alpha_1 = \alpha_2
\equiv \alpha = 0.2 \pm 0.05$. The numerical values of the other
parameters are standard and can be found in~\cite{Gatto:2010pt}.

\begin{figure}
\begin{center}
\includegraphics[width=7cm]{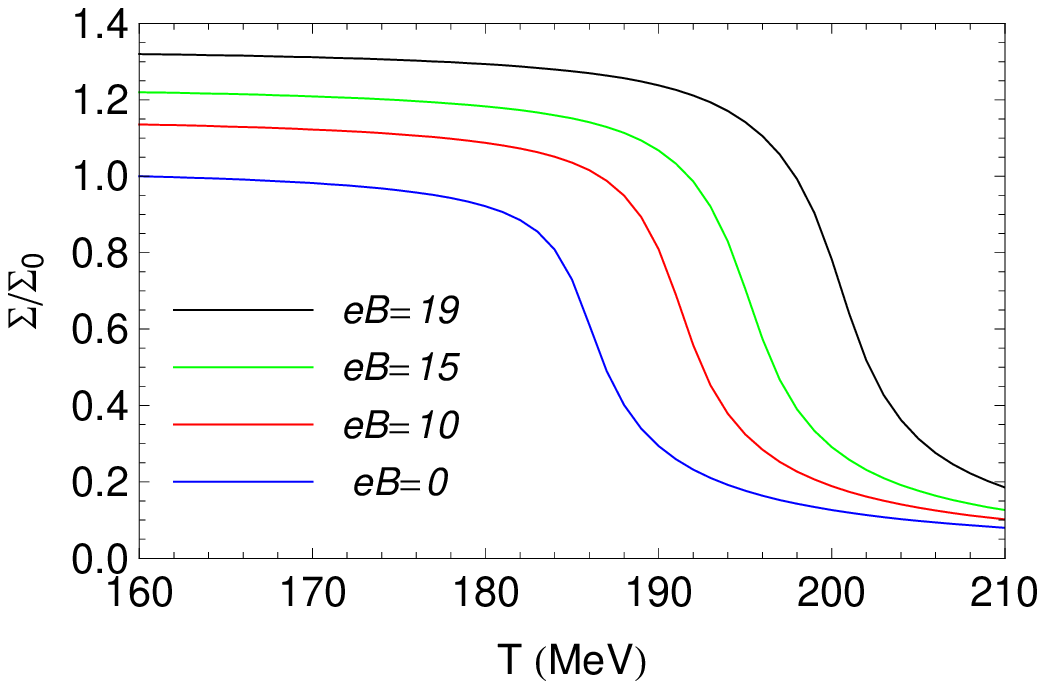}~~~\includegraphics[width=7cm]{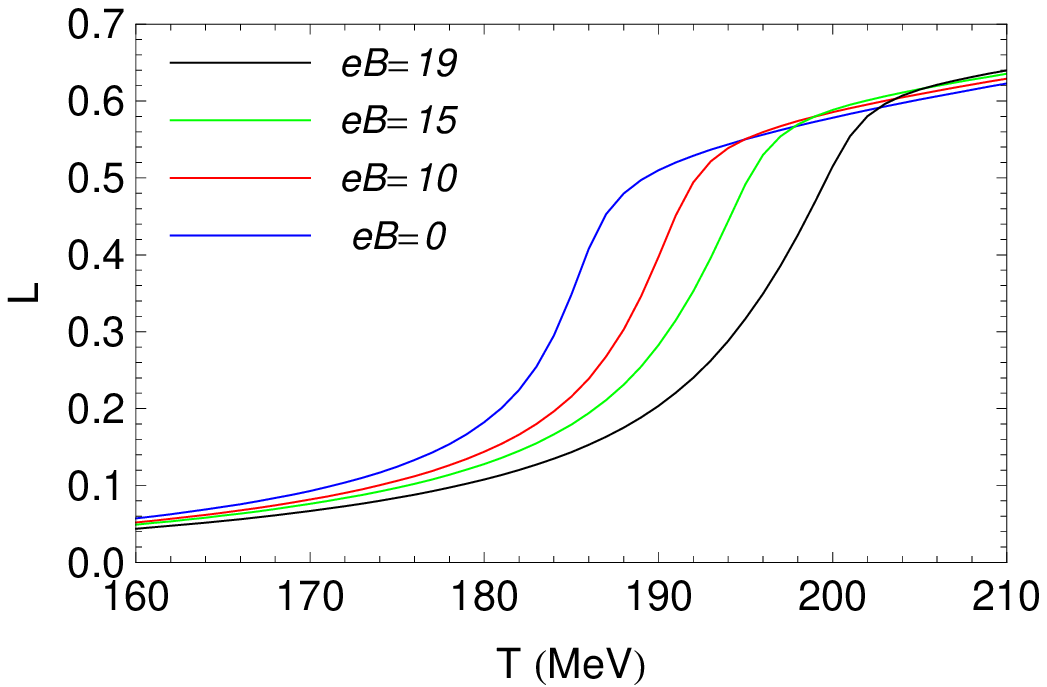}
\end{center}
\caption{\label{Fig:cond} {\em Left panel:} Chiral condensate
$\Sigma$, measured in units of the chiral condensate at zero
temperature and zero field, $\Sigma_0$, as a function of
temperature, for several values of the magnetic field strength.
{\em Right panel:} Expectation value of the Polyakov loop, $L$, as
a function of temperature, for several values of the magnetic
field strength. In the figures, the magnetic fields are measured
in units of $m_\pi^2$. Results correspond to the EPNJL model.}
\end{figure}

In Fig.~\ref{Fig:cond} I plot data for the chiral condensate
$\Sigma$, measured in units of the condensate at vanishing
temperature and magnetic field, that is $\Sigma_0 =
2\times(-253~\text{MeV})^3$, and the expectation value of the
Polyakov loop as a function of temperature, computed for several
values of the magnetic field strength. The most striking feature
of the results is that $T_L$ and $T_\chi$ are tied together also
in a strong magnetic field. At $\bm B=0$ the critical temperatures
are $T_P = T_\chi = 185.5$ MeV. Besides, at $eB=19 m_\pi^2$, $T_L
=199$ MeV and $T_\chi = 201$ MeV. This is due to the fact that the
NJL coupling constant in the pseudo-critical region in this model
decreases of the amount of the $15\%$ as a consequence of the
deconfinement crossover. Therefore, the strength of the
interaction responsible for the spontaneous chiral symmetry
breaking is strongly affected by the deconfinement, with the
obvious consequence that the numerical value of the chiral
condensate drops down and the chiral crossover takes place. The
picture remains qualitatively and quantitatively unchanged at
$eB=30 m_\pi^2$. In this case, $T_L = 224$ MeV and $T_\chi = 225$
MeV.

\section{The phase diagrams}
\begin{figure}
\begin{center}
\includegraphics[width=7cm]{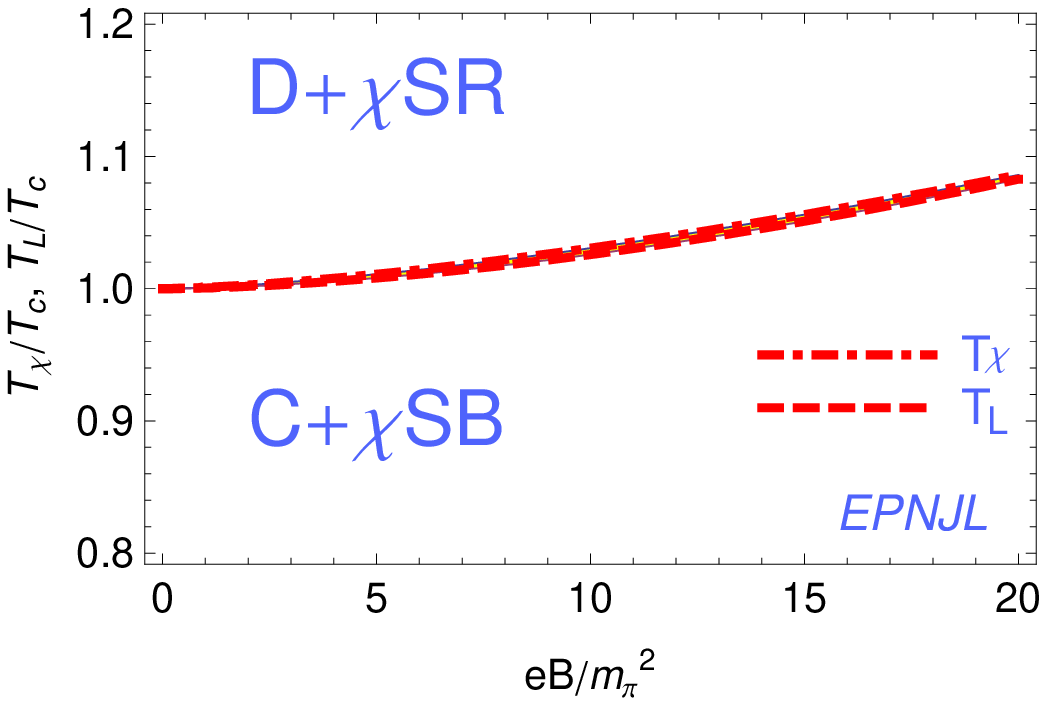}~~~\includegraphics[width=7cm]{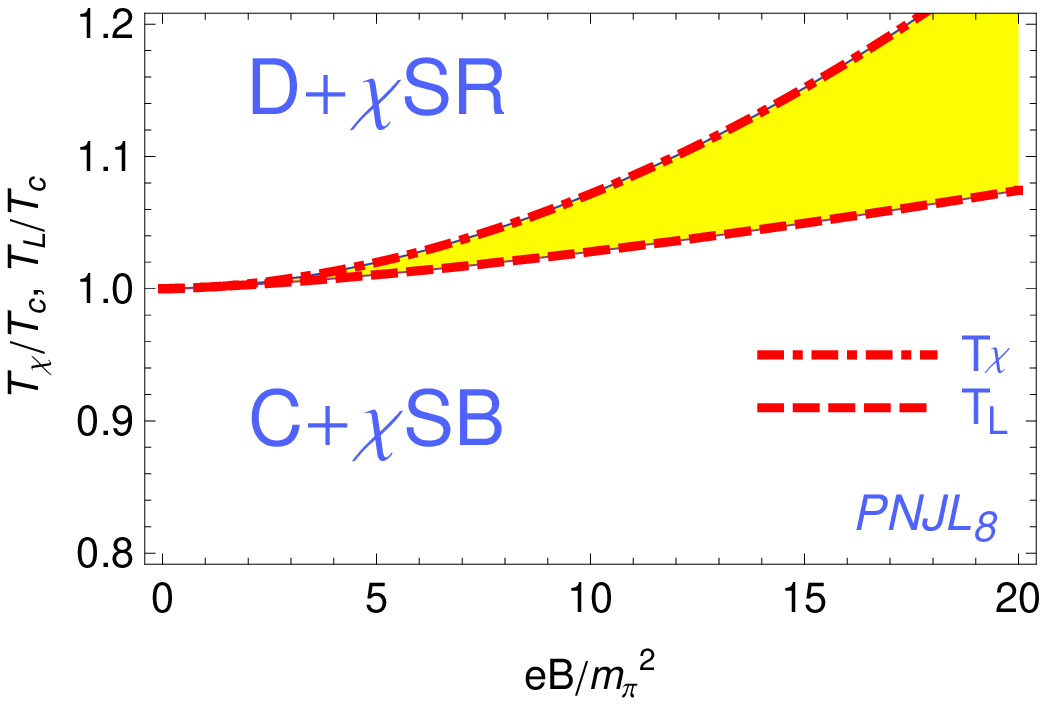}
\end{center}
\caption{\label{Fig:PD2} {\em Left panel:} Phase diagram in the
$eB-T$ plane for the EPNJL model. Temperatures on the vertical
axis are measured in units of the pseudo-critical temperature for
deconfinement at $eB=0$, namely $T_c = 185.5$ MeV. {\em Right
panel:} Phase diagram in the $eB-T$ plane for the PNJL$_8$ model.
Temperatures on the vertical axis are measured in units of the
pseudo-critical temperature for deconfinement at $eB=0$, namely
$T_c = 175$ MeV. In both the phase diagrams, $T_\chi$, $T_L$
correspond to the chiral and deconfinement pseudo-critical
temperatures, respectively. The golden region denotes the portion
of phase diagram in which hot quark matter is deconfined and
chiral symmetry is still broken spontaneously. }
\end{figure}

Data on deconfinement and chiral symmetry restoration
pseudo-critical temperatures are collected in Fig.~\ref{Fig:PD2},
in the form of a phase diagram in the $eB-T$ plane. The left panel
corresponds to the EPNJL model~\cite{Gatto:2010pt}; the right
panel corresponds to the results of the P-NJL$_8$
model~\cite{Gatto:2010qs}. In the figure, the magnetic field is
measured in units of $m_\pi^2$; temperature is measured in units
of the deconfinement pseudo-critical temperature at zero magnetic
field, namely $T_{B=0} = 185.5$ MeV for the EPNJL model, and
$T_{B=0} = 175$ MeV for the P-NJL$_8$ model. In the Figure, the
golden region corresponds to a phase in which hot quark matter is
deconfined, but chiral symmetry is still broken in a
non-perturbative way by the chiral condensate.

The most astonishing feature of the phase diagram of the P-NJL$_8$
model is the entity of the split among the deconfinement and the
chiral restoration crossover. The difference with the result of
the EPNJL model is that in the former, the entanglement with the
Polyakov loop is neglected in the NJL coupling constant. However,
the two models share one important feature: both deconfinement and
chiral symmetry restoration temperature are {\em increased} by a
strong magnetic field. They disagree quantitatively, in the sense
that the amount of split in the EPNJL model is more modest in
comparison with that obtained in the P-NJL$_8$ model. A similar
phase diagram calculation has been performed within the
Quark-Meson model in~\cite{Mizher:2010zb}, see also the talk by
Ana Mizher in this conference. In~\cite{Mizher:2010zb} it is shown
that the magnetic field leads to the split of deconfinement and
chiral symmetry restoration crossovers, if vacuum fermion
fluctuations are properly taken into account; furthermore, both
$T_\chi$ and $T_L$ are enhanced by a magnetic field, in excellent
agreement with the behavior of the NJL-like models.

The most recent Lattice data about this kind of study are those of
Ref.~\cite{D'Elia:2010nq}. In~\cite{D'Elia:2010nq}, the largest
value of magnetic field considered is $eB\approx 0.75$ GeV$^2$,
which corresponds to $eB/m_\pi^2 \approx 38$. The Lattice data
seem to point towards the phase diagram of the EPNJL model, since
no split is measured among the two crossovers. On the other hand,
the results of~\cite{D'Elia:2010nq} might not be definitive: the
lattice size might be enlarged, the lattice spacing could be taken
smaller (in~\cite{D'Elia:2010nq} the lattice spacing is $a=0.3$
fm), and the pion mass could be lowered to its physical value in
the vacuum. As a consequence, it will be interesting to compare
the model results with more refined data in the future.

\section{Conclusions and outlook}
In this talk, I have reported on recent
results~\cite{Fukushima:2010fe,Gatto:2010qs,Gatto:2010pt} about
the phase diagram of hot quark matter in a strong magnetic
background. I have considered two models: the P-NJL$_8$ model and
the EPNJL model, following the nomenclature
of~\cite{Sakai:2010rp}, which differ for the interactions content,
but both are tuned to fit Lattice data about QCD thermodynamics at
zero and imaginary chemical potential. I have also compared the
results with those of~\cite{Mizher:2010zb} in which a Quark-Meson
model is used. The most striking similarity among the several
models, when the contribution of the fermion determinant is
properly kept into account, is that they all support the scenario
in which chiral symmetry restoration and deconfinement
temperatures are {\em enhanced} by a strong magnetic field. The
models differ quantitatively for the amount of split measured
(very few percent for the EPNJL model, and of the order of $10\%$
for the other models for $eB \approx 20 m_\pi^2$). Furthermore, I
have also compared the model results with those obtained on the
Lattice~\cite{D'Elia:2010nq}.

It is fair to admit that the study presented here might have a
weak point, namely, it misses a microscopic computation of the
dependence of the NJL coupling on the Polyakov loop.
Equation~\eqref{eq:Run} is just a particular choice of such a
functional dependence. Different functional forms respecting the
$C$ and the extended $Z_3$ symmetry are certainly possible, and
without a rigorous derivation of Eq.~\eqref{eq:Run}, it merits to
study the topics discussed here using different choices for $G(L)$
in the near future. For these reasons, I prefer to adopt a
conservative point of view: it is interesting that the EPNJL
model, which is adjusted in order to reproduce Lattice data at
zero and imaginary chemical potential, predicts that the two QCD
transitions are entangled in a strong magnetic background;
however, this conclusion might not be definitive, since there
exist other model calculations which share a common basis with
EPNJL, and which show a more pronounced split of the QCD
transitions in a strong magnetic background. More refined Lattice
data will certainly help to discern which of the two scenarios is
the most favorable.

The results collected here suggest several continuation lines.
Firstly, it is worth to compute the magnetic susceptibility of the
chiral condensate~\cite{Ioffe:1983ju} within chiral models, which
allow to treat self-consistently the spontaneous chiral symmetry
breaking in a magnetic field. Besides, the technical machinery
used here can be easily applied to a microscopic study of the
spectral properties of mesons in strong magnetic field, in
particular of $\rho$ mesons following~\cite{Chernodub:2010qx}.
Furthermore, it would be interesting to make a complete analytical
study of the chiral limit, to estimate the effect of the magnetic
field on the universality class of two-flavor QCD. Finally, the
phase diagram with chiral chemical potential deserves further
study, following~\cite{Fukushima:2010fe,Chernodub:2011fr}.

\acknowledgments I am indebted to K.~Fukushima and R.~Gatto for
the valuable collaboration on which this talk is mainly based.
Besides, I acknowledge M.~Chernodub, E.~Fraga, M.~Frasca,
E.~M.~Ilgenfritz, T.~Kahara, K.~I.~Kondo, V.~Mathieu, A.~Mizher,
A.~Nedelin and H.~Warringa, for numerous discussions during our
stay in Gent. Besides, M.~D'Elia and F.~Negro are acknowledged for
discussions. Finally, it is a pleasure to acknowledge all the
organizers of the workshop for their kind invitation and warm
hospitality. The work of M.~R.\ is supported by JSPS under the
contract number P09028. The numerical calculations were carried
out on Altix3700 BX2 at YITP in Kyoto University.


\end{document}